\documentclass[aps,preprintnumbers,prl,twocolumn,superscriptaddress,nofootinbib,numbers,elide,compress]{revtex4-2} 
\usepackage{braket}
\usepackage{bbm,bm}
\usepackage{lmodern, amsmath, amssymb, slashed, graphicx, comment,adjustbox}
\usepackage[dvipsnames]{xcolor}
\usepackage{natbib}
\usepackage[colorlinks=true, citecolor=blue, urlcolor=blue, linkcolor=blue, breaklinks=true, pdfpagelabels=false]{hyperref}

\makeatletter\g@addto@macro\bfseries{\boldmath}\makeatother

\def\figureautorefname~#1\null{Fig.\,#1\null}
\def\equationautorefname~#1\null{Eq.\,(#1)\null}

\newcommand{\bpm}{\begin{pmatrix}}
\newcommand{\epm}{\end{pmatrix}}

\newcommand{\la}{\langle}
\newcommand{\ra}{\rangle}

\usepackage{pifont}

\DeclareMathOperator{\Tr}{Tr}

\newcommand{\bis}[1]{\begin{itemize}   \itemsep#1}
\newcommand{\bi}{\begin{itemize} }
\newcommand{\ei}{\end{itemize} }

\newcommand{\beq}{\begin{equation} }
\newcommand{\eeq}{\end{equation} }

\renewcommand{\sec}[1]{{{\textit{#1}}}.---}

\begin{document}

\title{Scattering entanglement entropy between one particle species and the others}
\author{Yuan-Hao Zheng}
\email{yuanhaozheng25@m.fudan.edu.cn}
\affiliation{Department of Physics and Center for Field Theory and Particle Physics,\\ Fudan University, Shanghai 200438, China}
\author{Jiayin Gu}
\email{jiayin\_gu@fudan.edu.cn}\thanks{corresponding author}
\affiliation{Department of Physics and Center for Field Theory and Particle Physics,\\ Fudan University, Shanghai 200438, China}
\affiliation{Key Laboratory of Nuclear Physics and Ion-beam Application (MOE), \\ Fudan University,
Shanghai 200433, China}

\begin{abstract}
We construct the entanglement entropy of a general scattering process between one particle species (denoted as $A$) and the rest of the final state.  Assuming a pure initial momentum eigenstate, we show that the reduced density matrix in particle $A$'s Fock space is block-diagonal in zero-, one- and multi-particle states, and is diagonal in the momentum basis for the one-particle state.  As a result, we could write down the Von Neumann entropy of particle $A$, $S_A$, in terms of scattering probabilities and differential cross sections, assuming no more than one particle $A$ is produced in the final state.  $S_A$ contains the information of both the total scattering probabilities (or cross sections) and the differential distributions --- the latter depends on a phase-space resolution factor as a result of the infinite-dimensional momentum Hilbert space.     
Our result is general and model-independent
since we consider the full final state from time evolution ($S|i\ra$) rather than a specific outcome, and is applicable as long as the probability for producing multiple particle $A$ is subleading.  We also obtain the $n\geq2$ Tsallis and R\'enyi entropies in our framework and find out that, at the leading order, they are proportional to the total scattering probability and carry no differential information.

\end{abstract}

\maketitle

\sec{Introduction} 
Scattering processes naturally generate quantum entanglements among the final-state particles.
Yet, our understanding of this entanglement --- how to quantify, calculate, and measure it --- remains incomplete, making it an active research frontier in high energy physics.
Theoretical studies have sought to define and compute the entanglement in scattering processes and to establish its relation to cross sections~\cite{Seki:2014cgq,Peschanski:2016hgk,Grignani:2016igg,Fan:2017hcd,Fan:2017mth,Schattschneider:2018zie,Peschanski:2019yah,Bose:2020shm,Fan:2021qfd,Fedida:2022izl,Cheung:2023hkq,Kowalska:2024kbs,Low:2024mrk,Low:2024hvn,MacIntyre:2025cmr,Sou:2025tyf,Fan:2025dum,Kowalska:2025qmf,Aguilar-Saavedra:2025njw,Peschanski:2026edo,Lyu:2025lja,Chang:2026nzq,Cheng:2025zaw}.
In parallel, rapid progress has been made in quantifying and measuring entanglement at high-energy colliders, particularly between the spin or flavor degrees of freedom of pair-produced particles~\cite{Belle:2007ocp,Afik:2020onf,Fabbrichesi:2021npl,Takubo:2021sdk,Gong:2021bcp,Severi:2021cnj,Afik:2022kwm,Aguilar-Saavedra:2022uye,Ashby-Pickering:2022umy,Aguilar-Saavedra:2022wam,Bi:2023uop,Fabbrichesi:2023cev,Han:2023fci,Ehataht:2023zzt,Barr:2024djo,ATLAS:2023fsd,CMS:2024pts,CMS:2024zkc,Cheng:2024btk,Cheng:2024rxi,Cheng:2025zcf,Aguilar-Saavedra:2022mpg,Fabbrichesi:2024rec,Fabbrichesi:2024wcd,Wu:2024asu,Altomonte:2024upf,Afik:2025grr,Cheng:2025cuv,Lin:2025eci,Han:2025ewp,Zhang:2025mmm,Aguilar-Saavedra:2025byk,Durupt:2025wuk,Guo:2026yhz,Fang:2026ddi,Hatta:2025obw,Fucilla:2026mkg,Joshi:2026zlb,Du:2026qsy,Pelliccioli:2026ltl,Goncalves:2026njf,Liu:2026dzv,deLima:2026jbb,Yang:2026rye,Liu:2026ees,Aoude:2026eeg,Cheng:2026ktp}.  
The role of entanglement in the partonic structure of hadrons has also been investigated, with proposed connections to final-state hadron multiplicities measured in deep inelastic scattering~\cite{Kovner:2015hga,Kharzeev:2017qzs,H1:2020zpd,Kharzeev:2021yyf,Hentschinski:2022rsa,Hentschinski:2023izh,Hentschinski:2024gaa,Hentschinski:2026otq}.
Entanglement with undetected radiation adds another dimension to this picture: tracing over unobserved radiation can induce decoherence in the observed system, an effect that is ubiquitous in realistic scattering events~\cite{Carney:2017jut,Carney:2017oxp,Gomez:2017rau,Carney:2018ygh,Neuenfeld:2018fdw,Tomaras:2019sjq,Irakleous:2021ggq,Burgess:2024heo,Aoude:2025ovu,Gu:2025ijz,Agrawal:2026zwa,Rey:2026yaa,Cheng:2026zfb}. 
Recent studies have also explored whether extremizing the entanglement in certain scattering processes can explain the values of Standard Model parameters~\cite{Cervera-Lierta:2017tdt,Quinta:2022sgq,Carena:2023vjc,Banerjee:2024lih,Thaler:2024anb,Garrido:2025xpk,Carena:2025wyh,Busoni:2025dns,Liu:2025bgw,Nunez:2025xds,Liu:2025iwh,Banerjee:2026pqw,Li:2026kha,Cao:2026aye,Dixit:2026yft,Yamagishi:2026ihf,Li:2026zum}. 
A central challenges in understanding the entanglement in scattering processes is related to the enormous degrees of freedom accessible to the final state, which spans an infinite-dimensional Fock space.  Focusing on specific scattering channels, such as elastic scattering, or selected particles or degrees of freedom can make the problem more tractable. A general framework that accommodates the full final state still remains desirable.

A scattering process is a unitary time evolution of the initial state $|i\ra$ with the $S$-matrix, given by  
\beq
S |i\ra = \sum_a |f_a \ra \la f_a| S |i\ra =  \sum_a \la f_a| S |i\ra |f_a \ra \,,
\eeq
where $|f_a\ra$ forms a complete basis for the final state.\footnote{For illustrative purposes, throughout the paper we will often use discrete labels even when the states are continuous (such as momentum).  It is implied that a continuum limit is taken, which is done explicitly later in the paper.} 
Typically, one focuses on a particular final state $|f_a\ra$ and calculate the corresponding matrix element $\la f_a| S |i\ra$. 
If $|i\ra$ is a pure state, so is $S|i\ra$.  The ``mixed-ness'' of a subsystem of the final state thus provides a measure of the entanglement between the subsystem and the rest of the final state.  This is naturally parameterized by the entanglement entropy, {\it i.e.} the von Neumann entropy of the subsystem, or generalizations of it.  

In this letter, we construct the entanglement entropy between one species of final-state particles (denoted as $A$) and the rest of the final state, assuming the initial state $|i\ra$ is a pure 4-momentum eigenstate.  
We show that, as a result of energy and momentum conservation, the reduced density matrix in particle $A$'s Fock space is block-diagonal in zero-, one- and multi-particle states, and is diagonal in the momentum basis for the one-particle state.  This makes it straightforward to write down the Von Neumann entropy of particle $A$, $S_A$, in terms of scattering probabilities and differential cross sections, at least up to one particle $A$ in the final state.


\vspace{0.2cm}
\sec{Entropy for one particle species} 
We start by decomposing the Fock space $\mathcal F$ of the final state as 
\beq
    \mathcal F=\mathcal F_A\otimes \mathcal F_{\rm others} \,,
\eeq
where the Fock space of particle $A$
\beq
\mathcal F_A=\ket0_A\oplus\mathcal H_A\oplus(\mathcal H_A\otimes\mathcal H_A)\oplus(\mathcal H_A\otimes\mathcal H_A\otimes\mathcal H_A)\oplus\cdots
\eeq 
may contain any number of particle $A$ (each with Hilbert space $\mathcal H_A$), including zero, which is simply denoted as $|0\ra_A$.  
$\mathcal F_{\rm{others}}$ is the Fock space of other particles. We can have a complete basis $\{\ket{\psi_i}_O\}$ in $\mathcal F_{\rm{others}}$ to define a partial trace,
\begin{align}
    \Tr_O(\cdots)=\sum_i \bra{\psi_i}_O\cdots\ket{\psi_i}_O.
\end{align}
One could then obtain the reduced density matrix of $A$, $\rho_A$, by applying $\Tr_O$ to the full density matrix of the final state.  
The entanglement entropy between $A$ and the rest of the outgoing state could then be constructed as the von Neumann entropy of the subsystem of $A$,
\beq
S_A = -\Tr(\rho_A \log \rho_A) \,.
\eeq
There always exists a basis in which $\rho_A$ is purely diagonal, and $S_A = -\sum_i p_i \log{p_i}$ where $p_i$ are the diagonal terms.  In general, such a basis could be difficult to find, and the basis states may not be eigenstates of particle numbers or momentum.  However, we will see that $S_A$ is block-diagonal in states with zero, one and multiple numbers of particle $A$,
\beq
\rho_A =
\bpm
p_0 & 0 & 0 \\
0 & p_1 \rho_{1A} & 0 \\
0 & 0 & p_{\rm mul} \rho_{{\rm mul}A}
\label{eq:blocked_density_matrix}
\epm \,,
\eeq
where $\rho_{1A}$ ($\rho_{{\rm mul}A}$) is the normalized reduced density matrix in the one- (multi-) particle subspace of $\mathcal F_A$. As such,
\begin{align}
S_A =&~  -p_0 \log p_0 - \Tr[p_1\rho_{1A} \log (p_1\rho_{1A} )] 
\nonumber\\ & ~~ - \Tr[p_{\rm mul}\rho_{{\rm mul}A} \log (p_{\rm mul}\rho_{{\rm mul}A} )] \nonumber \\
=&~ -p_0\log p_0-p_1\log p_1 +p_1 S_1 
\nonumber\\ & ~~ -p_{\rm mul}\log p_{\rm mul}+p_{\rm mul}S_{\rm mul}, \label{eq:sa00}
\end{align}
where 
\begin{align}
S_1 \equiv&~ -\Tr(\rho_{1A}\log \rho_{1A}) \,, \\
S_{\rm mul} \equiv&~ -\Tr(\rho_{{\rm mul}A}\log \rho_{{\rm mul}A}) \,
\label{eq:smul}
\end{align}
are the entanglement entropy in the corresponding one- and multi-particle subspaces.  Furthermore, $\rho_{1A}$ is automatically diagonal in the momentum basis.
As such, $S_1$ can be written in the form
\begin{align}
S_1
= - \sum_{\vec{k}, \alpha} p_{\vec{k},\alpha} \log p_{\vec{k},\alpha} \,, \label{eq:sa10}
\end{align}
where $\vec{k}$ is the 3-momentum of particle $A$ and $\alpha$ labels collectively any additional discrete degrees of freedom it has, such as spin, assuming it is in a basis for $\rho_A$ to be diagonal.\footnote{Throughout this letter, we use $k$ to denote momentum and reserve the lower-cased $p$ for probability.} Since $\rho_A$ in the $\alpha$ space may depend on the momentum, the diagonal basis of $\alpha$ itself may be momentum-dependent. Note that $\sum_{\vec{k}, \alpha} p_{\vec{k},\alpha} =1$ since we have normalized $\rho_{1A}$.


\vspace{0.2cm}
\sec{Prove of diagonality}
We will now prove that $\rho_A$ is block-diagonal in zero-, one- and multi-particle states and $\rho_{1A}$ is diagonal in momentum basis.  
Consider the evolution from an initial momentum eigenstate $\ket i$, with 4-momentum $P^\mu$. Due to energy and momentum conservation, we have the selection rule 
\begin{align}
    \bra{K^\mu;\alpha}S\ket i=0 \,, ~~\text{unless}~~ K^\mu=P^\mu,
\end{align}
where $K^\mu$ is the total 4-momentum of $\bra{K^\mu;\alpha}$, and $\alpha$ denotes collectively all other continuous or discrete labels it may have.  
The finial state could then be written as 
\begin{align}
    S\ket i=\sum_{k^\mu,\alpha,\beta} c_{k^\mu,\alpha,\beta} \ket{k^\mu;\alpha}_A \ket{P^\mu-k^\mu;\beta}_O \,,
\end{align}
where $\ket{k^\mu;\alpha}_A$ and $\ket{P^\mu-k^\mu;\beta}_O$ are states in $\mathcal{F}_A$ and $\mathcal{F}_{\rm others}$, respectively, with total momentum $k^\mu$ and $P^\mu-k^\mu$.
The total density matrix of the final state is
\begin{align}
\rho =&~ \sum
c_{k^\mu,\alpha,\beta}  c^*_{k'^\mu,\alpha',\beta'}  \nonumber\\  & ~~~ \ket{k^\mu;\alpha}_A \ket{P^\mu-k^\mu;\beta}_O \bra{k'^\mu;\alpha'}_A \bra{P^\mu-k'^\mu;\beta'}_O \,.
\end{align}
If we trace out $\mathcal{F}_{\rm others}$, only the terms with $\ket{P^\mu-k^\mu;\beta}_O = \ket{P^\mu-k'^\mu;\beta'}_O$ would survive. This imposes the condition $k^\mu = k'^\mu$, so $\rho_A$ is diagonal in the $k^\mu$ space,
\begin{align}
    \rho_A=\sum_{k^\mu,\alpha,\alpha'}  
    \rho_A(k^\mu, \alpha, \alpha') 
     \ket{k^\mu;\alpha}_A\bra{k^\mu;\alpha'}_A \,.
\end{align}
At this point, $\ket{k^\mu;\alpha}_A$ may contain any numbers of particle $A$, with $k^\mu$ their total 4-momentum.  However, the diagonalization in $k^\mu$ automatically separates zero-, one- and multi-particle states.  The zero-particle state, $\ket0_A$, is the only state with $k^\mu=0$\,.  A one-particle state always have $k^2 =m^2$ where $m$ is the mass of $A$ --- because of this, we can also replace the label $k^\mu$ by $\vec{k}$ since $k^0$ is not independent.  A multi-particle state always have $k^2 \geq 4m^2$.\footnote{A exception of this diagonalization is that if $A$ is massless, the one-particle state could not be separated from multiple collinear particles.  We do not address this problem here.}  On the other hand, multi-particle states with different particle numbers cannot be separated this way.  Note also that for one-particle momentum eigenstates, $\alpha$ contains only discrete degrees of freedom, such as spin and flavor, and $\rho_A$ can be diagonalized in the $\alpha$ space for a given $\vec{k}$.


\vspace{0.2cm}
\sec{An explicit expression for $S_A$}
We will now derive a more explicit expression of $S_A$ in the continuous $\vec{k}$ limit.  For simplicity, we will omit the multi-$A$ contribution, $-p_{\rm mul}\log p_{\rm mul}+p_{\rm mul}S_{\rm mul}$, which can always be restored in the end.  
In many cases, it is reasonable to assume the multi-$A$ contribution is suppressed,\footnote{For instance, if the center-of-mass energy is fixed and $A$ is sufficiently massive, the phase space for producing multiple $A$ is suppressed. In this case, we naturally get $p_{\rm mul} \ll p_0 , p_1$, while $S_{\rm mul}$ is also unlikely to be extremely large.} 
though there are important exceptions to it --- one of which is that the initial state contains more than one particle $A$, which we will address shortly. 
Combining \autoref{eq:sa00} and \autoref{eq:sa10}, we have
\begin{align}
S_A
=-p_0\log p_0-p_1\log p_1 - p_1 \sum_{\vec{k}, \alpha} p_{\vec{k},\alpha} \log p_{\vec{k},\alpha} \,.
\end{align}
Let us now take the continuum limit for $\vec{k}$, with
\beq
p_{\vec{k},\alpha} = p(\alpha|\vec{k}) p(\vec{k}) \to   p_\alpha(\Phi) \wp(\Phi) d \Phi \,,
\eeq
where $\Phi$ is the phase space of $\vec{k}$ and $\wp(\Phi)$ is the probability density in the $\Phi$ space, and $p_\alpha(\Phi)\equiv  p(\alpha|\vec{k})$ is the probability for particle A to have quantum number $\alpha$ for a given momentum $\vec{k}$. Here we have made it explicit that, for a given $\vec{k}$ (or $\Phi$), it is always possible to choose a basis for $\alpha$ such that $\rho_A$ is diagonal in the $\alpha$ space, with diagonal terms proportional to $p_\alpha(\Phi)$.
We choose the normalization such that 
\beq
\sum_\alpha p_\alpha (\Phi) = 1~\mbox{for any given}~\Phi \,, ~~ \int \wp (\Phi) d\Phi = 1 \,.
\eeq
In the following we will not explicitly write down the argument $\Phi$ of $p_\alpha (\Phi)$ and $\wp (\Phi)$.

We now have
\beq
S_A
= -p_0 \log p_0 -p_1\log p_1 - p_1 \sum_\alpha \! \int \! d\Phi \,  p_\alpha \wp  \log (p_\alpha \wp \Delta \Phi) \,,
\label{eq:sa0}
\eeq
where we have restored $d\Phi \to \Delta \Phi$ inside the log term as a regulator, or the phase-space resolution. 
As $\Delta \Phi \to 0$, $ \log \Delta\Phi$ diverges and $S_A$ is clearly ill-defined without a regulator. 
This dependence on the regulator comes from the fact that the von Neumann entropy typically scales with $\log N$ where $N$ is the dimension of the Hilbert space.  

Note that $S_A$ also contain the contribution of {\it no scattering}, that is, the initial-state particles pass through each other without any interactions.  If the initial state does not contain particle $A$, then this contribution is already included in $-p_0 \log p_0$. If the initial state contains one particle $A$, then $\wp(\Phi)$ contains a contribution $\sim p_{\rm ns} \delta^3 (\vec{k}-\vec{k}_{\rm initial})$, where the subscript ``ns'' stands for ``with $A$ in the initial state but no scattering.''  Since it occupies only a fixed point in the continuous $\Phi$ space, we can ignore its interference with the scattering amplitude and factorize out its contribution.  Schematically, this is just rewriting the $S_1$ in \autoref{eq:sa10} as  
\begin{align}
S_1
= - p_{\rm ns} \log p_{\rm ns} - \sum_{\vec{k}\neq \vec{k}_i , \, \alpha} p_{\vec{k}, \alpha} \log p_{\vec{k},\alpha} \,, \label{eq:sa11}
\end{align}
where $\vec{k}_i$ is the momentum of the particle $A$ in the initial state.   
Importantly,
here we have made an additional assumption that the initial particle $A$ is a pure momentum eigenstate and unentangled with the rest of the initial state. 
We can now write
\begin{align}
S_A =&~  - p_{\rm ns} \log p_{\rm ns} -p_0 \log p_0 -p_1 \log p_1 \nonumber\\ & ~~ - p_1 \sum_\alpha \int d\Phi \,  p_\alpha \wp  \log (p_\alpha \wp \Delta\Phi)  \,,
\label{eq:sa01}
\end{align}
where we have simply redefined $p_1$, $p_\alpha$ and $\wp $ such that they only include the scattered contributions.  
In particular, we now have $p_{\rm ns} + p_0 +p_1 (+p_{\rm mul}) =1$.  
Note that, if the initial state contains more than one particle $A$, similar arguments apply, and in most cases
we can redefine $p_{\rm mul}$ to move the no-scattering contribution to $p_{\rm ns}$.\footnote{Here we only consider the case where the initial state is $A+A$ with both $A$ being momentum eigenstates, unentangled with each other.  One subtle situation is that when the final state is also $A+A$, we cannot separate the no-scattering and scattered contributions in our framework since we consider all particle $A$ as one subsystem. Nevertheless, we can include it in the no-scattering contribution and \autoref{eq:sa01} still applies.   
}    
As such, we take \autoref{eq:sa01} to be general, with $p_{\rm ns}=0$ by definition if the initial state contains no particle $A$.\footnote{Again, if both the initial and the final state contain no particle $A$, we cannot separate the no-scattering and scattered contributions in our framework.  In this case $-p_0 \log p_0$ includes the no-scattering contribution, as already mentioned.}

In real collider experiments, scattering probabilities  are never directly measured --- instead, one extracts the cross section from measured event rates by normalizing to the luminosity, which accounts for the beam populations, collision frequency, and spatial overlap, {\it etc}.
Typically, the probability for a single pair of incoming particles to collide is extremely small. 
As such, one normally treats the incoming particles as plane waves, in which case the scattering probabilities 
formally tend to zero.\footnote{See {\it e.g.} discussions in Refs.~\cite{Low:2024mrk,Low:2024hvn}.  This is also the reason why a wave-packet formulation is adopted in these studies.} 
It is useful to consider this limit, where we have two possibilities:
\begin{itemize}

  \item $A$ is not in the initial state, and we simply have $ - p_{\rm ns} \log p_{\rm ns} =0$;

  \item $A$ is in the initial state, we have  
\beq   
  1- p_{\rm ns} = p_0 + p_1  \ll 1 \,, \nonumber
\eeq  
and 
\beq - p_{\rm ns} \log p_{\rm ns}  \approx  p_0 + p_1 \ll - p_0 \log p_0 - p_1  \log p_1 \,. \nonumber
\eeq

\end{itemize}
Either way, we can omit the $ - p_{\rm ns} \log p_{\rm ns}$ contribution and turn \autoref{eq:sa01}  back to \autoref{eq:sa0}, but with $p_1$, $p_\alpha$ and $\wp$ containing only scattered contributions.  Note that, following the same argument, we can also practically set $ - p_0 \log p_0 =0$ if the initial state contains no particle $A$.

With the no-scattering contribution removed from $\wp$, we could now write 
\beq
\wp = \frac{1}{\sigma} \frac{d\sigma}{d \Phi} \,,
\eeq
which is the normalized differential cross section in the phase space of particle $A$.  Note that $\sigma = \int d\sigma$ is the total cross section for producing exactly one particle $A$
together with any other possible final state particles (denoted collectively as $X$), 
\beq
\sigma \equiv \sigma ({\rm initial} \to A + X) \,.
\eeq
In general, it is not possible to separate the contributions from different states of $X$ in \autoref{eq:sa0}, since it is traced over.  However, if the phase spaces of different contributions have no overlap, the integration over $d\Phi$ can be automatically separated.  For instance, suppose the final state can be either $A+B$ or $A+C$, where particles $B$ and $C$ have different masses, their contributions are automatically separated since the particle $A$ have different values of $|\vec{k}|$ for the two cases.     

Let us also
modify the definition of $S_1$ to remove the no-scattering contribution so that we could restore the expression
\beq
S_A =
-p_0 \log p_0 -p_1 \log p_1 + p_1 S_1  \,,
\label{eq:sa1}
\eeq
where
\begin{align}
S_1 =&~  - \sum_\alpha  \int d\Phi \, p_\alpha \wp  \log (p_\alpha \wp \Delta \Phi) \nonumber \\
=&~ - \int d\Phi \wp  \log (\wp \Delta\Phi) - \int d\Phi \wp \, \sum_\alpha p_\alpha \log p_\alpha   \,.
\label{eq:sa1x}
\end{align}
We have broken $S_1$ into two terms. The first term, $- \int d\Phi \wp  \log (\wp \Delta\Phi)$, contains only momentum entanglement, and has an explicit dependence on the phase-space resolution $\Delta \Phi$.  The second term, $- \int d\Phi \wp \, \sum_\alpha p_\alpha \log p_\alpha$, contains information on the discrete degrees of freedom, and would be absent if we trace over the $\alpha$ space.  
Assuming $\wp$ is not a pure delta function ({\it i.e.} the distribution of $\vec{k}$ has some spread), for sufficiently small $\Delta \Phi$, the contribution from momentum entanglement to $S_1$ could be much larger than the ones from any discrete degrees of freedom, such as spin.  
This is simply because the former has much more degrees of freedom.   
We can also restore the multi-$A$ contribution, $-p_{\rm mul}\log p_{\rm mul}+p_{\rm mul}S_{\rm mul}$, in \autoref{eq:sa1}, but we do not have an explicit expression for $S_{\rm mul}$ like the one in \autoref{eq:sa1x} for $S_1$.

\vspace{0.2cm}
\sec{An example case} 
Consider the elastic scattering of two different scalar particles $A$ and $B$.    
It is possible to write a model and choose a center-of-mass energy such that $A + B \to A + B$ is the only possible process.  In this case, $p_0 = p_{\rm mul}=0$, we have
\beq
S_A =
-p_1 \log p_1 + p_1 S_1 \,,
\eeq
and
\beq
S_1 = - \int d\Omega \frac{1}{\sigma} \frac{d\sigma}{d\Omega} \log \left( \frac{1}{\sigma} \frac{d\sigma}{d\Omega}  \Delta \Omega \right)\,, 
\label{eq:sa1x1}
\eeq
where $d\Omega = d\cos \theta d\phi$ is the solid-angle element of particle $A$, and $\Delta \Omega$ the angular resolution. Note that $2\to2$ scattering is special in the sense that the magnitude of particle $A$'s momentum is fixed for a given center-of-mass energy, and the phase space is left with a 2-dimensional solid angle.  

The interpretation of \autoref{eq:sa1x1} is straightforward.  Schematically, subtracting the no-scattering contribution, the final state is
\beq
(S-1)|i\ra \sim \sum_{\theta,\phi} \mathcal{M}(\theta, \phi) |\theta, \phi \ra_A
|\pi-\theta, \phi+\pi \ra_B \,.
\eeq
The momenta (or angles) of particle $A$ and $B$ are thus generally entangled, and the more ``spread out'' $\frac{d \sigma}{d \Omega} \sim |\mathcal{M}(\theta, \phi) |^2$ is, the larger the entanglement. In particular, $S_1$ is maximized when $\frac{d\sigma}{d \Omega}$ is uniform in $\Omega$, corresponding to an equal partition of all states in $\mathcal{H}_A$, with
\beq
S_{1,\rm max} =  - \log \frac{\Delta \Omega}{4\pi} = \log N \,,
\eeq
where $N = \frac{4 \pi}{\Delta \Omega}$ is the total number of pixels on the solid angle.


\vspace{0.2cm}
\sec{Tsallis and R\'enyi entropies} 
We can also construct the Tsallis entropy~\cite{Tsallis:1987eu} for the example above, which is given by
\beq
T_n(\rho)=\frac{1-\Tr(\rho^n)}{n-1} \,,
\eeq
where the $n=2$ case is the so-called {\it linear entropy}.
In the Tsallis entropy, the no-scattering contribution cannot be omitted, but $p_{\rm ns}$ can still be factorized out as a diagonal term in $\rho_A$ (since $\rho_{1A}$ is diagonal in the momentum basis of $A$) as 
\beq
\rho_A = \bpm p_{\rm ns} &0 \\ 0 & p_1 \rho_{1A} \epm \,, \label{eq:rhotas}
\eeq
where $p_{\rm ns} = 1-p_1$, and the $p_1$ and $\rho_{1A}$ here contain only scattered contributions. For $n\geq2$, assuming a small $p_1$, we have
\begin{align}
T_n(\rho_A)=&~ \frac{1-(1-p_1)^n-p_1^n\Tr \rho^n_{1A} }{n-1} \nonumber \\
=&~ \frac{n}{n-1} p_1 + \mathcal{O}(p^2_1) \,. \label{eq:tsa}
\end{align}
This is in agreement with the results in Refs.~\cite{Low:2024mrk,Low:2024hvn} which use a wave-packet formulation, with the identification $p_1 = \frac{\sigma}{L^2}$, where $\sigma$ is the total cross section for $A+B\to A+B$ and $L^2$ is the transverse area of the wave packet.  On the other hand, the differential information contained in $\rho_{1A}$ only appears at $\mathcal{O}(p^n_1)$.  Note that \autoref{eq:tsa} generally holds for any final state with zero or one particle $A$ --- one only needs to replace $p_1$ by $p_0+p_1$ when the initial state contains particle $A$.\footnote{If particle $A$ is not in the initial state, one  simply replaces $p_{\rm ns}$ by $p_0 = 1-p_1$ and \autoref{eq:tsa} still holds. We do not discuss multi-$A$ production here.}

Similarly, one could construct the R\'enyi entropy~\cite{Renyi:1961entropy},
\beq
R_n (\rho) = \frac{\log \Tr \rho^n}{1-n} \,, 
\eeq
with von Neumann entropy being the $n\to 1$ limit.
For $n\geq2$, we obtain
\begin{align}
R_n (\rho_A) =&~ \frac{1}{1-n} \log \left [ (1-p_1)^n + p^n_1 \Tr \rho^n_{1A} \right] \nonumber \\
=&~ \frac{n}{n-1} p_1 + \mathcal{O}(p^2_1)  \,, 
\end{align}
which is the same as the Tsallis entropy at the leading order in $p_1$, as expected.  For $n\geq2$,
the Tsallis and R\'enyi entropies of particle $A$ are simply proportional to $p_1$ at the leading order in scattering probabilities.  Furthermore,
they are not sensitive to a small phase-space resolution $\Delta \Phi$ --- for finite $\wp$, $\Tr \rho^n_{1A} \sim \sum (\wp \Delta \Phi)^n$ simply vanishes in the $\Delta \Phi \to 0$ limit.  Both features are related to the fact that for $n\geq2$, the Tsallis and R\'enyi entropies enhance the contribution with the largest weight ($p_{\rm ns}$) and suppress the small ones (differential probabilities).


\vspace{0.2cm}
\sec{Summary and outlook} 
We have obtained, for the final state of a general scattering process, a formula for the von Neumann entanglement entropy between one particle species and the rest of the final state. 
As suggested by \autoref{eq:sa1}, this entropy $S_A$ factorizes into several pieces, 
and depends on the scattering probabilities and differential cross sections.  This relies on a key feature of the density matrix $\rho_A$ that it is block-diagonal in zero-, one- and multi-particle states, and diagonal in momentum basis for the one-particle state. 
The omitted contribution from multi-$A$ final state, $-p_{\rm mul}\log p_{\rm mul}+p_{\rm mul}S_{\rm mul}$, can also be restored in \autoref{eq:sa1}, but there is no simple expression for the multi-$A$ entropy $S_{\rm mul}$ that can be obtained within our framework.

The entanglement entropy we obtained depends on the extremely small, and practically unmeasurable scattering probabilities.  
This is exactly what one would expect, since scattering is what generates the entanglement, assuming the initial states are unentangled. The overall smallness of the entanglement entropy is merely a reflection that most initial state particles pass through each other without any interactions. 
On the other hand, the normalized entropy associated with the one-particle-$A$ subspace, $S_1$, is independent of the total scattering probability and contains the differential information in the channel $A+X$ (where $X$ contains any number of other particles).  
It inevitably contains a contribution $-\log\Delta \Phi$ from the phase-space resolution $\Delta \Phi$ due to the logarithmic  sensitivity of Von Neumann entropy to the dimension of the Hilbert space.  
This dependence on phase-space resolution or other regulators was also observed in previous studies~\cite{Grignani:2016igg,Carney:2017jut,Schattschneider:2018zie,Tomaras:2019sjq,Peschanski:2019yah,Bose:2020shm,Aguilar-Saavedra:2025njw,Rey:2026yaa}. For instance, a discretization of solid angles is performed in Ref.~\cite{Aguilar-Saavedra:2025njw} to obtain a well-defined momentum entanglement.  On the other hand, it might be possible to apply the wave-packet treatments in Refs.~\cite{Low:2024mrk,Low:2024hvn} to final-state particles, and the transverse area of the wave-packet could serve as a built-in resolution.  However, a wave-packet formulation may not be compatible with our approach, as the diagonality of $\rho_A$ relies on particles being momentum eigenstates.

We also calculated the $n\geq2$ Tsallis and R\'enyi entropies in our framework and found out that they are insensitive to the differential distribution at the leading order in the scattering probabilities.  
It would be interesting to further explore how different definitions of entropy carry different weights for the total rate and the differential information.  
It is also desirable to generalize our framework to consider any arbitrary bipartition of the final state.  Naively, it is straightforward to generalize the notion of particle $A$ to a group of particles.  An immediate challenge is that the $k^2$ for a group of particles is no-longer fixed, and in general the density matrix is not block-diagonal in one-group and multi-group states.  Another important question is that, when multiple particle $A$ are produced, whether one could still consider the subsystem with just one particle $A$ and obtain a meaningful expression for the entropy.  This might provide a way to tackle the multi-particle entropy $S_{\rm mul}$ (defined in \autoref{eq:smul}), which is not addressed in our framework.

\bigskip

{\noindent \bf Acknowledgement}~~We thank Ian Low, Ding Yu Shao, Lian-Tao Wang, Zhewei Yin and Yang Zhou for useful discussions and valuable comments on the manuscript.  This work is supported by the National Natural Science Foundation of China (NSFC) under Grant No.\,12375091. JG is also supported by the Innovation Program for Quantum Science and Technology under grant No.\,2024ZD0300101.

\bibliographystyle{JHEP}
\bibliography{references}

\end{document}